\documentclass[final,times,twocolumn,sort&compress]{elsarticle}
\usepackage{bm}
\usepackage{caption}
\usepackage{color}
\usepackage{epstopdf}
\usepackage{graphicx}
\usepackage{lmodern}
\usepackage{mathtools}
\usepackage{multirow}
\usepackage{makecell}
\usepackage{subfigure}
\usepackage{textcomp}
\usepackage{ulem}
\usepackage{url}

\begin{document}

\title{Interpreting the charge-dependent flow and constraining the chiral magnetic wave with event shape engineering}

\author[label1,label2]{Chun-Zheng Wang}
\author[label3]{Wen-Ya Wu}
\author[label3]{Qi-Ye Shou\corref{cor1}}
\ead{shouqiye@fudan.edu.cn}
\cortext[cor1]{Corresponding authors}
\author[label3]{Guo-Liang Ma}
\author[label3]{Yu-Gang Ma\corref{cor1}}
\ead{mayugang@fudan.edu.cn}
\author[label3]{Song Zhang}
\address[label1]{Shanghai Institute of Applied Physics, Chinese Academy of Sciences, Shanghai 201800, China}
\address[label2]{University of Chinese Academy of Sciences, Beijing 100049, China}
\address[label3]{Key Laboratory of Nuclear Physics and Ion-beam Application (MOE), Institute of Modern Physics, Fudan University, Shanghai 200433, China}

\begin{abstract}
The charge asymmetry ($A_{\rm ch}$) dependence of anisotropic flow serves as an important tool to search for the chiral magnetic wave (CMW) in heavy-ion collisions. However, the background effect, such as the local charge conservation (LCC) entwined with collective flow, has not yet been unambiguously eliminated in the measurement. With the help of two models, the AMPT with initial quadrupole moment and the blast wave (BW) incorporating LCC, we discuss the features of the LCC-induced and the CMW-induced correlations between $A_{\rm ch}$ and the flow. More importantly, we first propose to use the Event Shape Engineering (ESE) technique to distinguish the background and the signal for the CMW study. This method would be highly desirable in the experimental search for the CMW and provides more insights for understanding the charge-dependent collective motion of the quark-gluon plasma.
\end{abstract}

\maketitle

\section{Introduction} \label{sec:intro}

The interplay of the chiral anomaly and the ultra-strong magnetic field created in the off-central heavy-ion collisions could give rise to various anomalous chiral phenomena in the quark-gluon plasma (QGP)~\cite{Kharzeev:2008, Kharzeev:2016, Hattori:2017, Zhao:2019}, such as the chiral magnetic effect (CME), the chiral separation effect (CSE), the chiral magnetic wave (CMW)~\cite{Burnier:2011, Burnier:2012, Yee:2014, Taghavi:2015}, etc. The study of these novel phenomena is of fundamental significance since they may not only reveal the topological structure of vacuum gauge fields, but the possible local violation of $\cal P$ (parity) and/or $\cal CP$ (charge-parity) symmetries in strong interactions as well.

Over the past decade, the charge separations caused by the anomalous chiral effects have been sought by the STAR, ALICE and CMS collaborations at different collision energies and systems with multiple probes~\cite{Zhao:2019, Li:2020}. Though early measurements suggest some similarities between the observables and the theoretical expectations, there is considerable evidence that the background effects play a dominant role in the experimental measurement. In the CME study, for instance, the $\gamma$ correlator originally aiming at detecting the electric dipole moment proportionally varies as the elliptic flow ($v_2$) and the invariant mass of the particle pairs change~\cite{Wang:2018}, indicating the contribution from the local charge conservation (LCC) and/or transverse momentum conservation (TMC) entwined with collective flow. In recent years, substantial attempts are made to extract the fraction of the potential CME signal~\cite{STAR:2021, Tang:2020}. A general consensus is reached that the signal is consistent with zero within experimental uncertainties and the upper limit is no more than 20\% at 95\% CL in semi-central collisions~\cite{Zhao:2019}.

In the study of the CMW, the electric quadrupole moment is usually examined by the charge asymmetry ($A_{\rm ch}$) dependence of $v_2$ between the positively and negatively charged particles:
\begin{equation} \label{eq:1}
\Delta v_{2} \equiv v_{2}^{-} - v_{2}^{+} \simeq rA_{\rm ch}
\end{equation}
with $A_{\rm ch} \equiv (N^{+} -N^{-}) / (N^{+} +N^{-})$, 
or, equivalently, in the form of the covariance,
\begin{equation} \label{eq:2}
\langle v_{2}^{\pm} A_{\rm ch}\rangle-\langle A_{\rm ch}\rangle\langle v_{2}^{\pm}\rangle,
\end{equation}
which is also known as the three-particle correlator~\cite{ALICE:2016}. The linear relationship in Eq. (\ref{eq:1}) has been experimentally observed and the slope $r$ agrees well with theoretical prediction of the CMW~\cite{ALICE:2016, STAR:2015}. On the other hand, the non-CMW background also manifests itself in the identical relation in p-Pb collisions and for triangular flow ($v_3$)~\cite{CMS:2019}. Among several background sources~\cite{Bzdak:2013, Campbell:2013, Stephanov:2013, Hatta:2016, Hongo:2017, Zhao:2019a}, the LCC is believed to be the most prominent one. References~\cite{Bzdak:2013},~\cite{Voloshin:2014} and, recently,~\cite{Wu:2021} have clearly demonstrated that when selecting events with a specific $A_{\rm ch}$ value, one preferentially applies nonuniform kinematic cuts on charged particles, leading to the $A_{\rm ch}$-dependent correlation. Though the existence of the LCC is a known fact, few works, however, have directly addressed the central issue of the CMW measurement: how to disentangle the LCC background from the possible signal. In the CME study, the Event Shape Engineering (ESE) technique~\cite{Schukraft:2013, Wen:2018}, which investigates the observable as a function of $v_2$, has proved to be an effective way to estimate the flow-related background and constrain the magnitude of the signal. Unfortunately, similar approach has not yet been established for the study of CMW and the strength (or the fraction) of the signal remains unexplored.

In this Letter, we discuss and compare the features of the CMW-induced and the LCC-induced $A_{\rm ch}$-$v_2$ correlations with the help of two models, the AMPT with initial quadrupole moment and the blast wave (BW) incorporating LCC. Based on that, we then propose, for the first time, a ESE method to differentiate the LCC background and to extract the CMW fraction, which would be highly desirable and feasible in experimental search for the CMW.

\section{Model description} \label{sec:mod}

The hybrid transport model AMPT is widely used in simulating relativistic heavy-ion collisions. The string melting version, in particular, is known for describing the collective motion of the final state hadrons~\cite{Lin:2005}. AMPT consists of four subroutines which simulate different stages of the collisions in sequence: HIJING for the initial parton condition~\cite{Wang:1991}, ZPC for the partonic evolution~\cite{Zhang:1998}, a simple quark coalescence for the hadronization process and ART for the hadronic rescatterings and interactions. In this work, the parameters of Lund string fragmentation and the cross section of parton scattering are set to reproduce the hadron spectrum and anisotropic flow at RHIC and LHC energies~\cite{Lin:2014, Xu:2011}. To mimic the CMW-induced electric quadrupole moment, the approach proposed in Refs.~\cite{Ma:2011, Ma:2014} is performed, which interchanges the $y$ component of the position coordinate for some in-plane light quarks carrying positive (negative) charges with those out-plane ones carrying negative (positive) charges. The strength of the quadrupole moment is determined by the number of quarks being interchanged. Previous studies suggest that switching $\approx$3\% of total quarks can generate a comparable slope with the experimental result in semi-central collisions~\cite{Ma:2014, Shen:2019}.

The Monte Carlo droplet generator DRAGON~\cite{Tomasik:2009} describes an expanding and locally thermalized fireball, which decays into fragments and subsequently emit hadrons. Phase space distribution of the fragments is based on the BW model~\cite{Retiere:2004}, which assumes that the radial expansion velocity is proportional to the distance from the centre of the system. The elliptic shape of the fireball is controlled by a geometry parameter and the elliptic flow can be further generated by another parameter $\rho_2$ in form of $\rho_2 \cos(2\phi)$ with $\phi$ being the boost angle. The LCC effect is additionally incorporated by forcing charged particles to emit $always$ in pairs with zero net charge (one positively and one negatively charged) at the same spatial point~\cite{ALICE:2020}. The momenta of particles in each pair are independently sampled and then boosted together so particles eventually follow a common collective velocity given by the single-particle BW configuration. Note that such a procedure maximizes the LCC effect since all particles are pair-produced. A recent study suggests a smaller LCC fraction ($\approx$1/3) in RHIC energy~\cite{STAR:2021}. Our goal here is to qualitatively present the feature of the LCC so the parameter is not precisely tuned. More realistic treatment should strictly take into account the multiplicity and the charge balance function, which is worth a try in future work. For simplicity, all particles are set to have pion mass.

One should be aware that AMPT model barely has the LCC effect at final stage since the parton rescattering and the coalescence procedure have largely distorted initial spatial charge distribution in HIJING~\cite{Du:2007, Wu:2021}. Therefore, the AMPT and the BW simulate only the CMW signal and the LCC background respectively. In this work, we sample both models roughly in semi-central collisions without accurately determining the centrality.

\section{Charge asymmetry dependence of elliptic flow} \label{sec:v2ach}

\begin{figure}
\centering
\includegraphics[width=\linewidth]{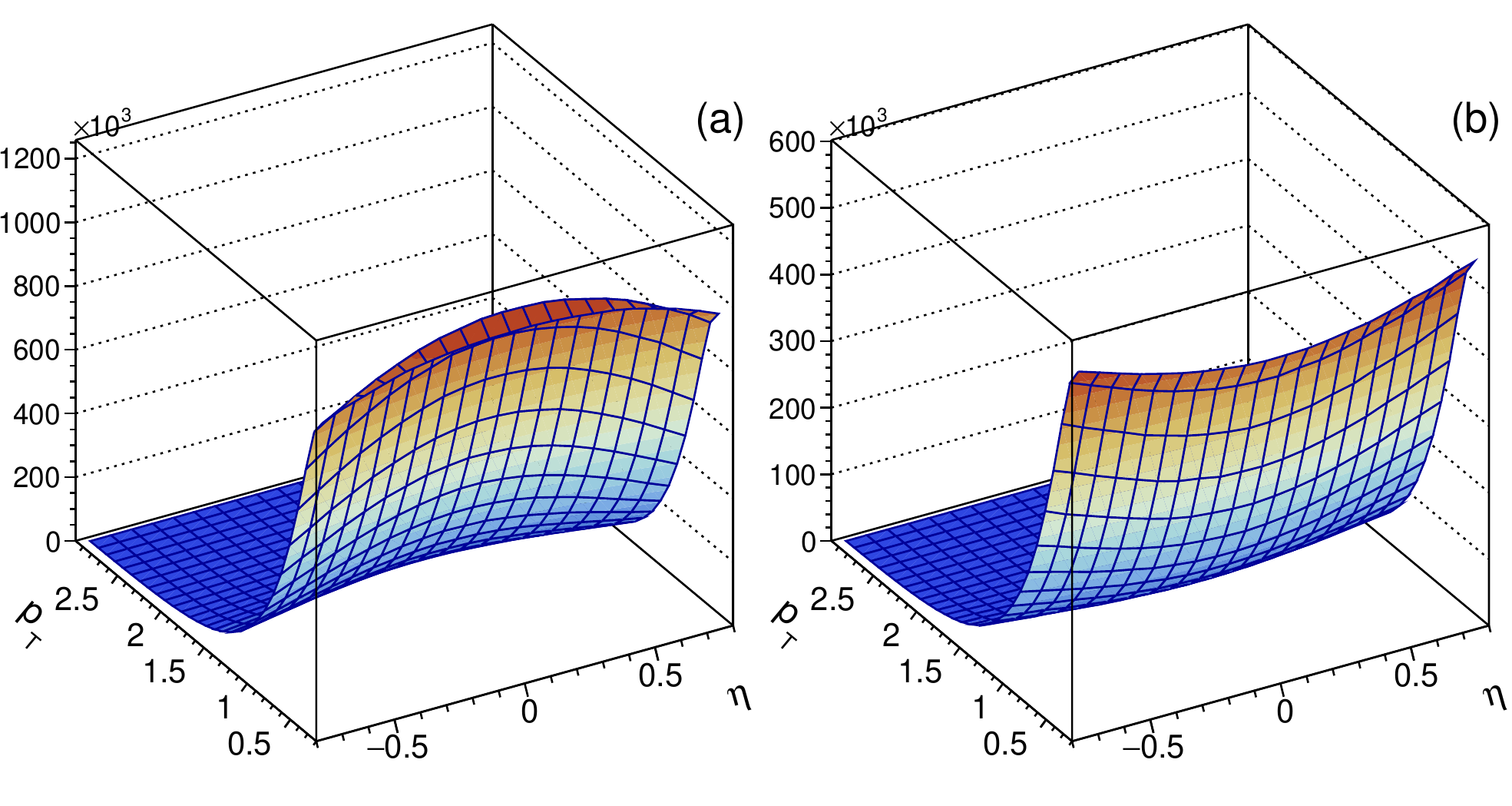}
\captionsetup{justification=raggedright}
\captionof{figure}{2-D histogram of $p_{\rm T}$ and $\eta$ when (a) both particles in the pair are detected and (b) only one particle in the pair is detected in the BW+LCC model.}
\label{fig:dist_pt_eta}
\end{figure}

The dependence of $v_2$ on $A_{\rm ch}$ in AMPT has been studied in Refs.~\cite{Ma:2014, Shen:2019, Xu:2020}. Without introducing initial quadrupole moment, AMPT fails to reproduce the $A_{\rm ch}$-$v_2$ relation and the contribution from the resonance decay can be either negative or positive depending on the mass. In the presence of the charge separation, however, the linear dependence can be perfectly formed, demonstrating the applicability of the observable.

\begin{figure}
\centering
\includegraphics[width=\linewidth]{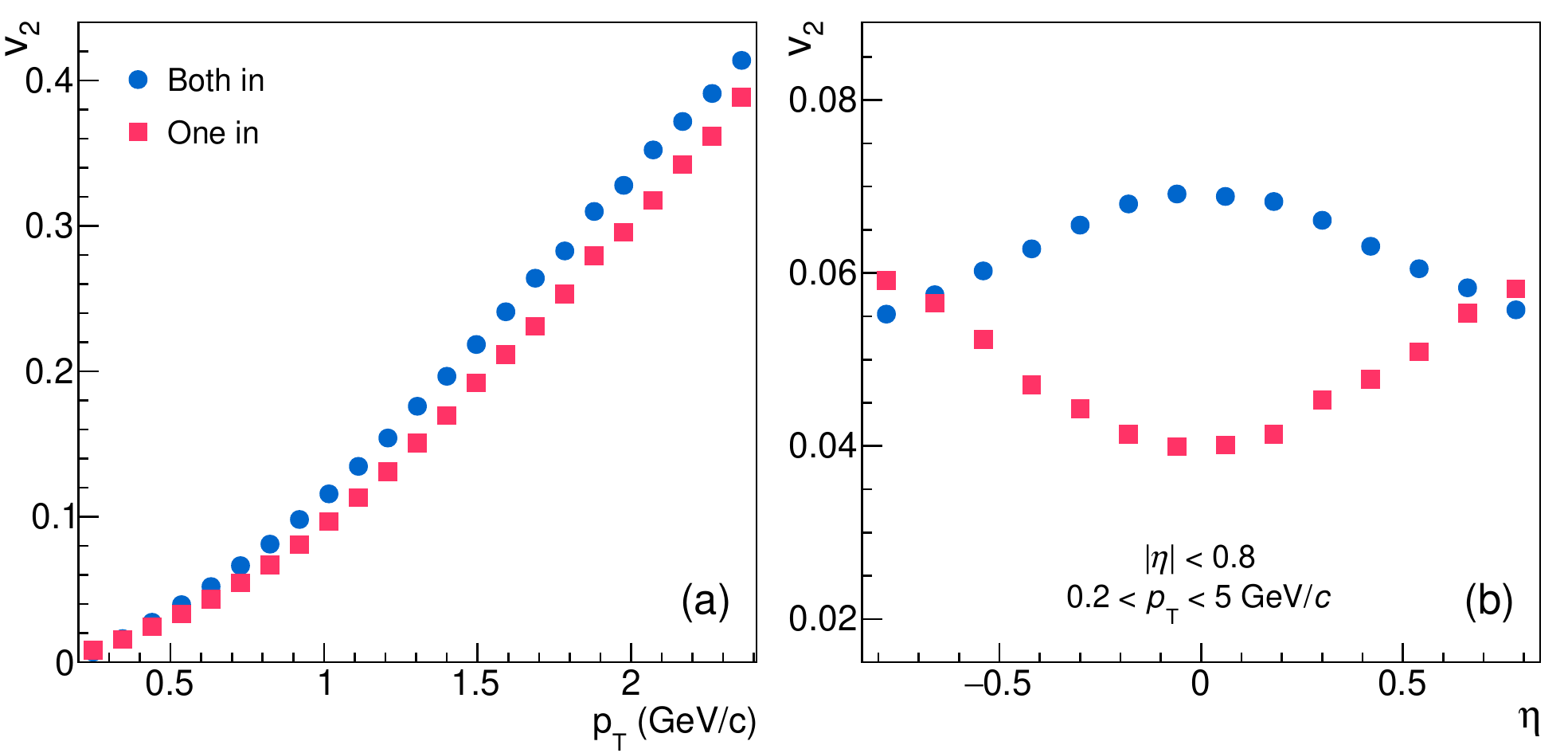}
\captionsetup{justification=raggedright}
\captionof{figure}{$v_2$ as a function of (a) $p_{\rm T}$ and (b) $\eta$ for the cases of ``both in" and ``one in" in the BW+LCC model.}
\label{fig:v2_pt_eta}
\end{figure}

Now we focus on the $A_{\rm ch}$-dependent correlation in BW+LCC with $\rho_2$ = 0.3. Figure~\ref{fig:dist_pt_eta} presents the 2-D histogram of $p_{\rm T}$ and $\eta$ for the case that (a) both particles in the pair are within the detector acceptance, denoted as ``both in", and (b) only one particle in the pair is within the detector, denoted as ``one in". It can be seen that particles carry higher (lower) $p_{\rm T}$ and/or smaller (larger) $|\eta|$ tend to be (un)paired, contributing to (non)zero $A_{\rm ch}$. This picture agrees well with the mechanism proposed in Refs.~\cite{Bzdak:2013, Wu:2021}: selecting events with a specific $A_{\rm ch}$ is, essentially, cutting on particles with nonuniform kinematic windows. The $p_{\rm T}$ and $\eta$ distributions of particles in events with different $A_{\rm ch}$ values are completely different. Therefore, the $v_2$, which remarkably depends on $p_{\rm T}$ and $\eta$, also varies with $A_{\rm ch}$. Figure~\ref{fig:v2_pt_eta} shows the $v_2$ as a function of (a) $p_{\rm T}$ and (b) $\eta$ for the above two cases. The $v_2$($p_{\rm T}$) for both cases are quite similar despite that the value of ``both in" is slightly larger than the one of ``one in" over all $p_{\rm T}$ ranges. On the other hand, the $v_2$($\eta$) for two cases significantly differ from each other. The $v_2$ value of ``both in" at $\eta \simeq 0$ is 1.5-2 times larger than that of ``one in". Such a big discrepancy of $v_2$ directly comes from the $\eta$ distribution shown in Fig.~\ref{fig:dist_pt_eta}: a clear convex shape for ``both in" and a concave shape for ``one in" in the $\eta$ direction. A typical event with nonzero $A_{\rm ch}$ is dominated by the ``one in" case. Consequently, as more unpaired particles with a given charge are detected, the lower the average $v_2$ of such particles is. Figure~\ref{fig:v2_pt_eta} also explains the reason why $A_{\rm ch}$-$p_{\rm T}$ cannot interpret the $A_{\rm ch}$-$v_2$ slope alone, which puzzles the measurement for years~\cite{Shou:2019}. The LCC background cannot be entirely eliminated by simply narrowing down the $p_{\rm T}$ coverage since the differential distributions of $v_2$($p_{\rm T}$) and $v_2$($\eta$) need to be comprehensively taken into account. 

\begin{figure}
\centering
\includegraphics[width=\linewidth]{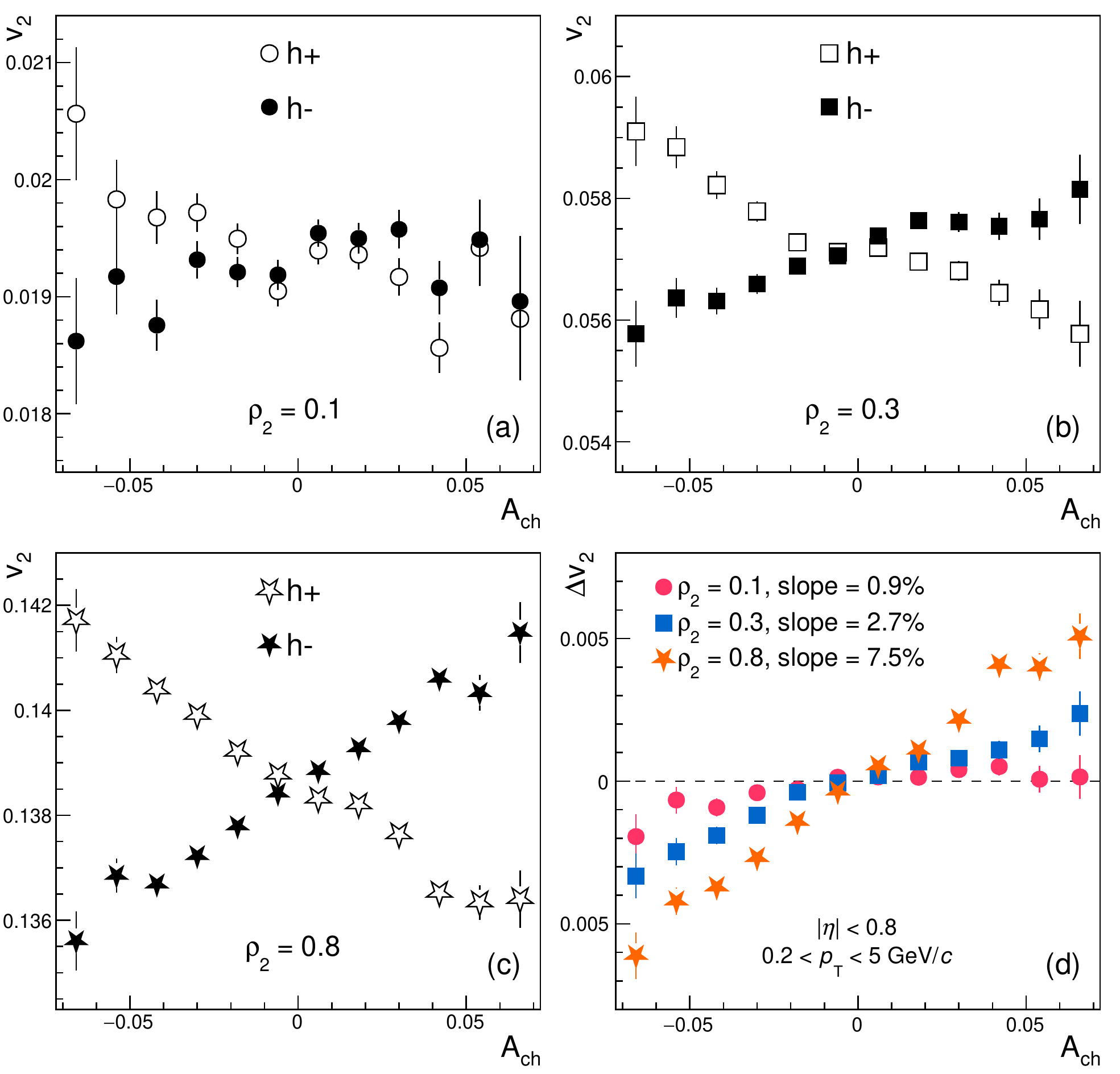}
\captionsetup{justification=raggedright}
\captionof{figure}{(a-c) The $A_{\rm ch}$-$v_2$ correlations with varied $\rho_2$ values. (d) The slopes of $A_{\rm ch}$-$\Delta v_2$ are compared between different $\rho_2$.}
\label{fig:v2_ach}
\end{figure}

The $A_{\rm ch}$-$v_2$ correlations are examined with varied $v_2$ values as shown in Fig.~\ref{fig:v2_ach} (a)-(c). The initial elliptic geometry is fixed in our model and the $v_2$ is only tuned by the parameter $\rho_2$. The selected $\rho_2$ values are 0.1, 0.3 and 0.8, corresponding to the integrated $v_2$ of 0.02, 0.06 and 0.14 at midrapidity region, respectively. The slopes of $A_{\rm ch}$-$\Delta v_2$ are compared in Fig.~\ref{fig:v2_ach} (d). It can be clearly seen that the larger the $\rho_2$ is, the stronger the slope is. Such a behavior reveals an important feature of the LCC background: the magnitude of the slope is positively associated with the $v_2$. It is also confirmed in this study that after normalizing the $\Delta v_2$ by the sum ($v_2^+ + v_2^-$), all slopes are in line with each other, therefore the similarity between the normalized slopes reported in experiments~\cite{CMS:2019, Shou:2019} implies the dominance of the LCC effect. Considering that the CMW-induced slope, in contrast, weakly correlates with $v_2$, the $v_2$-dependent slope can be used to distinguish the CMW signal from the LCC background, which will be further discussed in Sec.~\ref{sec:ese}. 

\section{Three particle correlation} \label{sec:3pc}

In addition to the slope of $A_{\rm ch}$-$v_2$, the three particle correlator is another noteworthy observable. The integral three particle correlator calculates the covariance of $A_{\rm ch}$ and $v_2$ as defined in Eq. (\ref{eq:2}). One advantage of measuring such a covariance instead of the slope is that the former is free of the correction for efficiency of $A_{\rm ch}$ in the experiment~\cite{Voloshin:2014, ALICE:2016}. The differential three correlator, which measures the correlation between the flow at a particular kinematic space and the charge of the third particle (rather than the eventwise $A_{\rm ch}$) at another particular coordinate, reads,
\begin{equation}
\langle v_{n}^{\pm} q_{3} \rangle - \langle q_{3} \rangle \langle v_{n} \rangle,
\end{equation}
where the subscript $3$ denotes the third particle. This observable is usually examined as a function of the separation of pseudorapidity ($\Delta \eta$) between the first and the third particle, resembling the study of the charge balance function. The experimental results of the differential correlator has been reported by the ALICE collaboration~\cite{ALICE:2016}. A nontrivial dependence of the correlator on $\Delta \eta$ is observed, roughly matching the expectation of the LCC background~\cite{Voloshin:2014}. Unfortunately, as also mentioned in~\cite{ALICE:2016}, no prediction was made from the perspective of the CMW. The AVFD framework has not yet been able to achieve it~\cite{Shi:2018}. The AMPT with initial quadrupole provides an option for the phenomenological estimation.

\begin{figure}
\centering
\includegraphics[width=\linewidth]{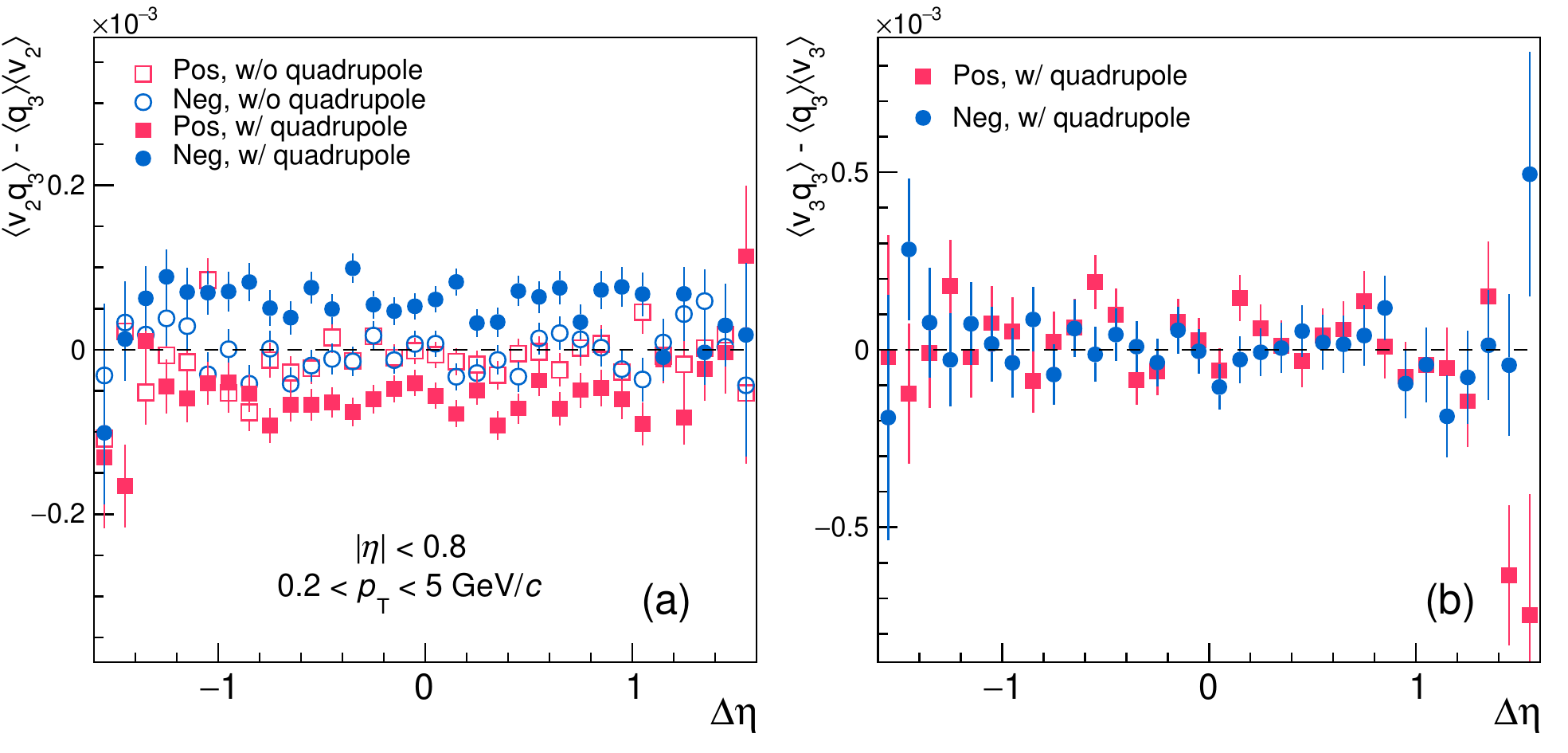}
\captionsetup{justification=raggedright}
\captionof{figure}{The differential three particle correlator for the (a) second and the (b) third harmonics as a function of $\Delta \eta$.}
\label{fig:cov}
\end{figure}

Figure~\ref{fig:cov} presents the differential three particle correlator for the (a) second and the (b) third harmonics as a function of $\Delta \eta$. Without initial quadrupole, the second order correlator is consistent with zero regardless of the charge of the first particle, which agrees with the zero slope of $A_{\rm ch}$-$v_2$ observed in original AMPT~\cite{Ma:2014}. In the presence of the quadrupole moment, a clear separation between the correlators can be seen. When the flow particle is positively (negatively) charged, the third particle surrounded tend to be negative (positive), generating the negative (positive) covariance. This trend qualitatively matches the ALICE measurement~\cite{ALICE:2016}. Nevertheless, no $\Delta \eta$ dependence of the differential correlator is found in our study since the initial charge separation is implemented uniformly in the $\eta$ direction. At large $\Delta \eta$, the correlators in this work remain constant while the experimental ones remarkably decrease and change the sign. Besides, the third order correlator in the ALICE data exhibits a similar structure as the second harmonic despite the smaller strength while it is consistent with zero in our model. Such differences indicate that the ALICE measurement of the differential correlator is very likely to be dominated by the LCC mechanism rather than the CMW.

\section{Event shape engineering} \label{sec:ese}

As presented in Sec.~\ref{sec:v2ach}, the slope of the LCC-induced $A_{\rm ch}$-$v_2$ roughly scales with the magnitude of $v_2$. Therefore, it is viable to disentangle the LCC background contributions from the possible CMW signal by investigating the observable at different $v_2$ values. The Event Shape Engineering (ESE) technique fortunately meets the needs. In a given centrality interval, the ESE method is able to select events with different initial geometry fluctuations based on the flow vector:
\begin{equation} \label{eq:4}
q_{n}=\frac{|Q_{n}|}{\sqrt{M}},
\end{equation}
where $M$ is the multiplicity, and $Q_{n}$ is defined as
\begin{equation} \label{eq:5}
Q_{n, x}=\sum_{i=1}^{M}\cos(n\phi_{i}), \quad Q_{n, y}=\sum_{i=1}^{M}\sin(n\phi_{i}),
\end{equation}
with $\phi_{i}$ being the azimuthal angle of the $i$-th particle. This method has been widely used in the study of CME~\cite{CMS:2019} and other topics concerning the collectivity~\cite{ALICE:2016b}. Here we divide the whole AMPT sample into ten $q_2$ bins according to Eq. (\ref{eq:4}) and only present top five bins due to the statistics. For the BW model, two approaches are adopted: (1) sample the $\rho_2$ value event-by-event from a uniform distribution and then apply the ESE to the mixed data, (2) instead of performing the ESE, simply set the $\rho_2$ value to be 0.1, 0.3, 0.8 and 1.2, and then calculate $v_2$ and observables for each case. The latter treatment takes account of the fact that, unlike AMPT, the $v_2$ fluctuations in BW model is rather small.

\begin{figure}
\centering
\includegraphics[width=\linewidth]{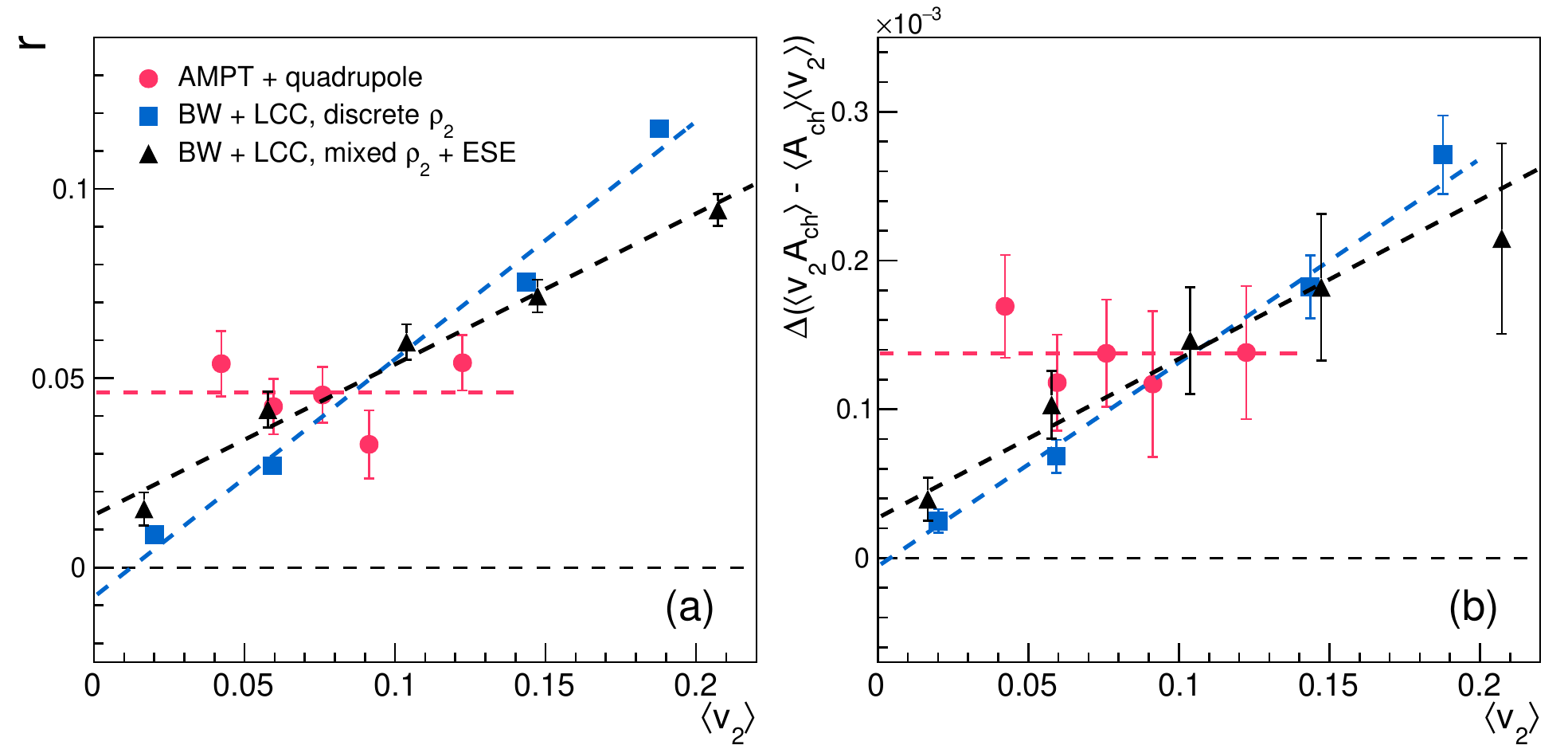}
\captionsetup{justification=raggedright}
\captionof{figure}{(a) The slope of $A_{\rm ch}$-$\Delta v_2$ and (b) the difference of the integral three particle correlator as a function of average $v_2$.}
\label{fig:ese}
\end{figure}

Figure~\ref{fig:ese} shows (a) the slope of $A_{\rm ch}$-$\Delta v_2$ and (b) the difference of the integral three particle correlator as a function of average $v_2$. Experimentally, the later observable has the advantage in statistics since it does not require dividing the data sample into several $A_{\rm ch}$ bins. The results are compared between two models. For both observables, the results of AMPT with quadrupole do not exhibit any $\langle v_2 \rangle$ dependence. The observables remain unchanged even if $\langle v_2 \rangle$ is reduced by half. A linear fit gives a significantly positive intercept at zero $\langle v_2 \rangle$, indicating the strength of the CMW signal. In contrast, the results of BW with LCC are found to be proportional to $\langle v_2 \rangle$. The observables linearly decrease as $\langle v_2 \rangle$ decreases and the intercept at zero $\langle v_2 \rangle$ for the discrete $\rho_2$ method is consistent with zero. Note that the intercept for the mixed  $\rho_2$ method is slightly about zero, which is a natural fluctuation coming from the initial distribution of $\rho_2$ used to sample the data. The reason that the signal and the background response differently to the ESE lies in their disparate origins: the former stems from the quadrupole configuration regardless of $v_2$ while the latter, as demonstrated in Sec.~\ref{sec:v2ach}, is an effect purely from manipulating $v_2$ with kinematic windows.

What we present here are two ideal extremes: the model contains either CMW or LCC alone. Realistically, the observables measured in the experiment are very likely to include both the signal and the background. In that case, the data points may form a linear relationship with a positive slope and a nonzero intercept. The CMW fraction, or more accurately, the upper limit can then be extracted by the ratio between the observable at zero $v_2$ and at finite $v_2$:
\begin{equation} \label{eq:6}
f_{\rm CMW} = \frac{b}{a\langle v_2 \rangle+b},
\end{equation}
where $a$ and $b$ are the slope and the intercept from the linear fit. This method has actually been successfully implemented to constrain the CME fraction~\cite{CMS:2019}, so we believe that extending it to the CMW study should be highly feasible. At last, we remind experimentalists that, when performing this ESE approach, the factor between the slope of $A_{\rm ch}$-$\Delta v_2$ and the integral three particle correlator as mentioned in Ref.~\cite{ALICE:2016}, as well as the $A_{\rm ch}$ distribution at different $q_2$ intervals, need to be carefully checked since they may notably influence the fitting result of Eq. (\ref{eq:6}).

\section{Summary} \label{sec:sum}

The $A_{\rm ch}$-dependent flow, serving as the most probable probe for the search for the CMW, is investigated by two models, the AMPT with initial quadrupole moment and the BW incorporating LCC, which simulate the CMW signal and the LCC background respectively. In the BW+LCC scenario, we confirm that the $A_{\rm ch}$-$v_2$ relation can stem from the intrinsic property of $A_{\rm ch}$ as suggested in our previous study~\cite{Wu:2021}. It is revealed that the differential $v_2$, particularly $\eta$-dependent $v_2$, between the clusters contributing to zero and nonzero $A_{\rm ch}$ are significantly different, which naturally gives rise to the $A_{\rm ch}$-$v_2$ relation. More importantly, the slope or the covariance between $A_{\rm ch}$ and $v_2$ generated by the LCC mechanism is found to be proportional to the event $v_2$. This key feature makes it feasible to disentangle the LCC background from the CMW signal since the CMW-induced $A_{\rm ch}$-$v_2$ correlation does not exhibit strong $v_2$ dependence according to the simulation of AMPT+quadrupole. We propose, for the first time, the ESE method to estimate the strength of the LCC background and to extract the CMW fraction, which would be highly desirable and doable in experimental search for the CMW. In addition, the differential three particle correlator is also studied by AMPT with and without initial charge separation, which serves as a baseline to interpret the experimental measurement.

\section*{Acknowledgement}

We are grateful to S. A. Voloshin and G. Wang for enlightening discussions. We also thank W.-B. He and C. Zhong for their assistance. This work is supported by the Strategic Priority Research Program of Chinese Academy of Sciences (No. XDB34030000), the National Natural Science Foundation of China (Nos. 11890710, 11890714, 11890711, 11975078, 12061141008, 11421505, 11605070) and the National Key Research and Development Program of China (Nos. 2016YFE0100900, 2018YFGH000173). Q.S. is sponsored by the Shanghai Rising-Star Program (20QA1401500).

{}

\end{document}